\documentclass[prd,twocolumn,floats,floatfix,showpacs,nofootinbib]{revtex4}
\usepackage{graphicx}
\usepackage{dcolumn}
\usepackage{bm}
\usepackage{graphics}
\usepackage{slashed}
\usepackage{amssymb}
\usepackage{natbib}
\usepackage{amsmath}
\usepackage{amsthm}
\usepackage{url}
\usepackage{color}
\newcommand{\fracc}[2]{\frac{\textstyle{#1}}{\textstyle{#2}}}

\begin{document}

\title{Gordon Metric Revisited}

\author{M. Novello\footnote{M. Novello is Cesare Lattes ICRANet
Professor}}\email{novello@cbpf.br}
\author{E. Bittencourt}\email{eduhsb@cbpf.br}

\affiliation{Instituto de Cosmologia Relatividade Astrofisica ICRA -
CBPF\\ Rua Dr. Xavier Sigaud, 150, CEP 22290-180, Rio de Janeiro,
Brazil}

\pacs{04.20.-q}
\date{\today}

\begin{abstract}
We show that Gordon metric belongs to a larger class of geometries, which are responsible to describe the paths of accelerated bodies in moving dielectrics as geodesics in a metric $\hat q_{\mu\nu}$ different from the background one. This map depends only on the background metric and on the motion of the bodies under consideration. As a consequence, this method describes a more general property that concerns the elimination of any kind of force acting on bodies by a suitable change of the substratum metric.
\end{abstract}

\maketitle

\section{Introduction}

In 1923 Gordon \cite{gordon} made a seminal suggestion to treat the propagation of electromagnetic waves in a moving dielectric, modifying the metric structure of the background. He showed that the electromagnetic waves propagate as geodesics not in the background geometry $ \eta_{\mu\nu} $ but instead in the effective metric
\begin{equation}
\hat g^{\mu\nu} = \eta^{\mu\nu} + ( \epsilon \mu - 1 ) \,
v^{\mu} \,v^{\nu}, \label{251211}
\end{equation}
where $ \epsilon$ and $ \mu$ are constant parameters that characterize the dielectric and $v^{\mu}$ is the four-velocity of the material under consideration (which is not necessarily constant). Later, it was recognized that this interpretation could be used to describe nonlinear structures even when $\epsilon$ and $\mu $ depend on the intensity of the electromagnetic field \cite{novello1} or more complicated functions of the field strengths \cite{pendry}. In all these cases, the causal cone, which is associated to the effective metric, does not coincide with the null-cone of the theory. The origin of this modification is due to the presence of a moving dielectric, which changes the paths of the electromagnetic waves inside this medium.

We then face the question: could such particular description of the electromagnetic waves in moving dielectrics be generalized for other cases, in which accelerated paths due to other kind of forces would be described as geodesic motions in an associated metric? We shall see that the answer is affirmative and this kinematical map depends only upon the acceleration of the body and the background metric.

This method allows us to geometrize any force in the sense that arbitrary accelerated body in a given metric substratum $g_{\mu\nu}$ is equivalently described as geodesic motion in an effective geometry $\hat q_{\mu\nu}$. We start by analyzing the generalization of Gordon metric concerning the propagation of electromagnetic waves inside arbitrary dielectric media. This procedure mimics the trajectory which led to the geometrization of the gravitational field as it was done by general relativity (GR). This means to describe the effects of acceleration of a particle on a gravitational field by a change of the space-time metric, according to Einstein's approach. We compare this effective geometry to the metric in the post-Newtonian approximation in order to see if it is possible to reproduce some results of GR. Such procedure is exactly what happens in the analog models of gravitation that deals with systems kinematically equivalence, but dynamically distinct.

\section{Geometrizing accelerated paths}

Consider a vector field\footnote{The natural velocity field we use to develop this section is a given $1-$form field $u_{\mu}$. In Gordon approach, the wave vector $k_{\mu}$ is a gradient and, therefore, an exact $1-$form.} $u_{\mu}$ with norm $N\equiv u^{\alpha}u_{\alpha}$ in a given background $g_{\mu\nu}$. The acceleration of $u_{\mu}$ is given by
$$a_{\mu}=u_{\mu;\nu}\,u^{\nu}.$$
Now let us construct an associated metric tensor $\hat q_{\mu\nu}$, in which the vector $u_{\mu}$ satisfies the equation

\begin{equation}
\label{geo_gen1}
u_{\mu||\nu}\,\hat u^{\nu}=f(p)\,u_{\mu},
\end{equation}
where $||$ means covariant derivative with respect to $\hat q_{\mu\nu}$ and $f(p)$ is an arbitrary function of the parameter $p$ along the curve. The contravariant components of the vector field are defined by $\hat u^{\mu}\equiv\hat q^{\mu\nu}u_{\nu}$ and, consequently, the norm defined in $\hat q_{\mu\nu}$ is given by $\hat N\equiv\hat q^{\mu\nu}u_{\mu}u_{\nu}$. Whenever $u_{\mu}$ is either a gradient or a normalized vector field, $f(p)$ can be set equal to zero without loss of generality.

Developing Eq.\ (\ref{geo_gen1}), we obtain

\begin{equation}
\label{geo_gen}
\fracc{1}{2}\hat N_{,\mu}+u_{[\mu,\nu]}\hat u^{\nu}=f(p)\,u_{\mu},
\end{equation}
where $[\,]$ means skew-symmetrization. Choosing $\hat q_{\mu\nu}$ such that this equation is verified, then the accelerated path of $u_{\alpha}$ in $g_{\mu\nu}$ becomes a geodesic motion in the associated $\hat q_{\mu\nu}$. Note that this is independent of its functional form. We will show that Gordon metric is a particular example of this procedure and that there is a class of metrics which play the same role in such ``unforced-motion" process exhibiting different geometrical properties for each element of the class.

\section{Light paths on moving dielectric: Gordon approach}
Let us define two skew-symmetric tensors $F_{\mu\nu}$ and $P_{\mu\nu}$ representing the electromagnetic field inside the material medium. These tensors are expressed in terms of the field strengths $E^{\mu}$ and $H^{\nu}$ and field excitations $D^{\mu}$ and $B^{\mu}$ as follows

\begin{equation}
\nonumber
\begin{array}{lcl}
F_{\mu\nu} & \equiv & E_{\mu}v_{\nu} - E_{\nu}v_{\mu} + \eta_{\mu\nu}{}^{\alpha\beta}v_{\alpha}B_{\beta},\\[2ex]
P_{\mu\nu} & \equiv & D_{\mu}v_{\nu} - D_{\nu}v_{\mu} + \eta_{\mu\nu}{}^{\alpha\beta}v_{\alpha}H_{\beta},
\end{array}
\end{equation}
where $v^{\mu}$ is a given four-vector comoving with the dielectric and $\eta_{\mu\nu}{}^{\alpha\beta}$ is the Levi-Civita tensor. We assume that the electromagnetic properties of the medium are characterized by the constitutive relations

$$D_{\alpha}=\epsilon_{\alpha}{}^{\nu}(E,H)E_{\nu}, \hspace{.5cm} B_{\alpha}=\mu_{\alpha}{}^{\nu}(E,H)H_{\nu},$$
where $\epsilon_{\alpha}{}^{\nu}(E,H)$ and $\mu_{\alpha}{}^{\nu}(E,H)$ are arbitrary tensors. Consider Maxwell equations on dielectric media \cite{landau} with permittivity $\epsilon$ and permeability $\mu$ that characterize the dielectric:

\begin{equation}
\label{max_diel}
\begin{array}{l}
P^{\mu\nu}{}_{;\nu}=0,\\[2ex]
^*F^{\mu\nu}{}_{;\nu}=0.
\end{array}
\end{equation}
From now on, we take the background metric as flat Minkowski space-time and assume that $\mu\equiv\mu_0$ is a constant and $\epsilon=\epsilon(E)$, where $E\equiv\sqrt{-E_{\alpha}E^{\alpha}}$ and $E^{\alpha}$ is the electric field. It is straightforward to generalize these equations to arbitrary curved space-time. Indeed, suppose an observer with velocity $v^{\mu}$ comoving with the dielectric and such that $v^{\mu}{}_{;\nu}=0$. Then, Eqs.\ (\ref{max_diel}) written in terms of the displacement vectors $D^{\mu}$ and $B^{\mu}$ become

\begin{equation}
\begin{array}{l}
D^{\mu}{}_{;\nu}v^{\nu}-D^{\nu}{}_{;\nu}v^{\mu}+\eta^{\mu\nu\alpha\beta}v_{\alpha}H_{\beta;\nu}=0,\\[2ex]
B^{\mu}{}_{;\nu}v^{\nu}-B^{\nu}{}_{;\nu}v^{\mu}-\eta^{\mu\nu\alpha\beta}v_{\alpha}E_{\beta;\nu}=0.
\end{array}
\end{equation}

The projection with respect to $v^{\mu}$ yields the four independent non-linear equations of motion describing the electromagnetic field inside the dielectric medium:

\begin{equation}
\begin{array}{l}
\label{max_diel_proj}
\epsilon E^{\alpha}{}_{;\alpha}-\fracc{\epsilon'E^{\alpha}E^{\beta}}{E}E_{\alpha;\beta}=0,\\[2ex]
\mu_0H^{\alpha}{}_{;\alpha}=0,\\[2ex]
\epsilon \dot E^{\lambda} - \fracc{\epsilon'E^{\lambda}v^{\alpha}E^{\mu}}{E}E_{\mu;\alpha} + \eta^{\lambda\beta\rho\sigma}v_{\rho}H_{\sigma;\beta}=0,\\[2ex]
\mu_0 \dot H^{\lambda}-\eta^{\lambda\beta\rho\sigma}v_{\rho}E_{\sigma;\beta}=0.
\end{array}
\end{equation}
We define the unitary vector $l^{\mu}$ by setting $E^{\mu}\equiv El^{\mu}$, where $l^{\mu}$ satisfies $l_{\alpha}l^{\alpha}=-1$.

We use Hadamard conditions \cite{hadamard} to obtain the propagation waves through the characteristics surface $\Sigma$ (for details, see appendix). The symbol ${[X]}_{\Sigma}$ represents the discontinuity of $X$ through this surface. Then, the discontinuities of Eqs.\ (\ref{max_diel_proj}) become

\begin{equation}
{[E_{\mu,\lambda}]}_{\Sigma}=e_{\mu}k_{\lambda}, \hspace{1cm} {[H_{\mu,\lambda}]}_{\Sigma}=h_{\mu}k_{\lambda},
\end{equation}
where $e_{\mu}(x)$ and $h_{\mu}(x)$ are the amplitudes of the discontinuities and $k_{\mu}\equiv\partial_{\mu}\Sigma$ is the wave vector. Thus, it follows that

\begin{equation}
\begin{array}{l}
\epsilon k^{\alpha}e_{\alpha}-\fracc{\epsilon'}{E}E^{\alpha}e_{\alpha}E^{\beta}k_{\beta}=0,\\[2ex]
\mu_0h^{\alpha}k_{\alpha}=0,\\[2ex]
\epsilon k^{\alpha}v_{\alpha}e^{\mu} - \fracc{\epsilon'}{E}E^{\lambda}e_{\lambda}v^{\alpha}k_{\alpha}E^{\mu} + \eta^{\mu\nu\alpha\beta}k_{\nu}v_{\alpha}h_{\beta}=0,\\[2ex]
\mu_0 k_{\alpha}v^{\alpha}h^{\lambda}-\eta^{\lambda\beta\rho\sigma}k_{\beta}v_{\rho}e_{\sigma}=0,
\end{array}
\end{equation}
where $\epsilon'$ is the derivative of $\epsilon$ with respect to $E$. Combining these equations we obtain the following intermediary relation

\begin{equation}
\begin{array}{l}
\fracc{e^{\mu}}{\mu_0 k_{\alpha}v^{\alpha}} [k^{\nu}k_{\nu} - (k^{\nu}v_{\nu})^2] - \fracc{k^{\beta}e_{\beta}}{\mu_0 k_{\alpha}v^{\alpha}}k^{\mu}+\epsilon k^{\alpha}v_{\alpha}e^{\mu} +\\[2ex]
- \fracc{\epsilon'}{E}E^{\lambda}e_{\lambda}v^{\alpha}k_{\alpha}E^{\mu}=0,
\end{array}
\end{equation}
which multiplying by $E_{\mu}$ yields the dispersion relation

\begin{equation}
\left(\eta^{\mu\nu}+(\mu_0\epsilon-1+\mu_0\epsilon'E)v^{\mu}v^{\nu}-\fracc{\epsilon'}{\epsilon E}E^{\mu}E^{\nu}\right)k_{\mu}k_{\nu}=0.
\end{equation}
We see that the envelop of discontinuity propagates differently from Minkowski light-cone of the linear Maxwell theory. In this case, the causal structure is given by an effective Riemannian geometry\footnote{Mathematically, the metric tensor is a covariant tensor of rank 2. However, in this paper, we sometimes shall call ``metric" a contravariant tensor of rank 2. In particular, that is the way the Gordon metric appears naturally.} $\hat g^{\mu\nu}$. From this point of view, $k_{\mu}$ is null-like in $\hat g^{\mu\nu}$, namely,
\begin{equation}
\hat g^{\mu\nu}k_{\mu}k_{\nu}=0.
\end{equation}
The expression of the effective geometry is given by

\begin{equation}
\hat g^{\mu\nu} = \eta^{\mu\nu} + (\mu_0\epsilon-1+\mu_0\epsilon'E)v^{\mu}v^{\nu}-\fracc{\epsilon'E}{\epsilon}l^{\mu}l^{\nu}.
\end{equation}
A simple calculation show that its inverse is

\begin{equation}
\hat g_{\mu\nu} = \eta_{\mu\nu} - \left( 1-\fracc{1}{ \mu_0\epsilon(1+\xi)}\right)v_{\mu}v_{\nu}+\fracc{\xi}{1+\xi}l_{\mu}l_{\nu},
\end{equation}
where
\begin{equation}
\nonumber
\xi\equiv\fracc{\epsilon'E}{\epsilon}.
\end{equation}

In particular, when $\epsilon$ is a constant, this formula reduces to Gordon pioneer work, in which was shown that the waves propagate as geodesics not in the background geometry $\eta_{\mu\nu}$ but instead in the effective metric

\begin{equation}
\hat g^{\mu\nu} = \eta^{\mu\nu} + (\epsilon\mu_0 - 1) \,
v^{\mu} \,v^{\nu},
\end{equation}
which depends only on the dielectric properties $\mu_0$, $\epsilon$ and $v^{\mu}$. The magnitude $N$ of the wave vector in Minkowski space-time (written in terms of dielectric properties) is determined by Gordon relation

\begin{equation}
\label{mod_gord_k}
\begin{array}{l}
\hat g^{\mu\nu}k_{\mu}k_{\nu} = \left(\eta^{\mu\nu} + (\epsilon\mu_0 - 1) \,
v^{\mu}v^{\nu}\right)k_{\mu}k_{\nu}=0,\Longrightarrow\\[2ex]
\Longrightarrow N=(1-\mu_0\epsilon)(k.v)^2,
\end{array}
\end{equation}
where $k.v\equiv k_{\alpha}v^{\alpha}$.

The analysis of the wave propagation in material media and the study of effective geometry are particularly interesting in the investigation of analog model \cite{novello1,novello2} for the understanding of kinematical properties at very small scale of astrophysical objects (see details in \cite{novello3,novello4}). We quote Hawking radiation \cite{hawk_rad} and Unruh's work on experimental black hole evaporation \cite{unruh} which are systematically studied and the modeling of specific dielectric media is developed in order to eventually detect these tiny effects. A more complete discussion on this topic was given in \cite{nov_viss_volo,barcelo} and references therein. Here we shall point the similarity between these geometries in a later section.

\section{Binomial metrics}

In recent years an intense activity concerning features of Riemannian geometries similar to those described by Gordon approach has been done \cite{novello2}. In particular, those that allows a binomial form for both the metric and its inverse. That is, its covariant and the corresponding contravariant expressions are
\begin{equation}
\widehat{q}_{\mu\nu} = A\, \eta_{\mu\nu} + B \, \Phi_{\mu\nu},
\label{2612112}
\end{equation}
and
\begin{equation}
\widehat{q}^{\mu\nu} =  \alpha\,\eta^{\mu\nu} + \beta \, \Phi^{\mu\nu}.
\label{2612111}
\end{equation}
This form of the metric requires that the tensor $  \Phi^{\mu \nu} $ must satisfy the condition
\begin{equation}
\Phi_{\mu \nu} \,   \Phi^{\nu \lambda} \, = m \,   \delta_{\mu}^{\lambda} + n \,   \Phi_{\mu}^{\lambda}.
\label{2}
\end{equation}
Such feature allows us to write the inverse metric similarly to the binomial form of the metric, avoiding difficulties with infinite series\footnote{This is the case of general relativity as a field theory formulation. The exact expression for metric tensor is set
$$g^{\mu\nu} \equiv \eta^{\mu\nu}+ h^{\mu\nu}.$$
A consequence is that its inverse, the covariant tensor $g_{\mu\nu}$ is an
infinite series:
$$ g_{\mu\nu} = \eta_{\mu\nu} -  h_{\mu\nu} + h_{\mu\alpha} \, h^{\alpha}{}_{\nu} - h_{\mu\alpha} \, h^{\alpha}{}_{\beta}\, h^{\beta}{}_{\nu}+...$$ This formulation was introduced by Feynman, Gupta and others (cf. \cite{feyn}).}. Two remarkable examples of this property are the scalar field (in which $\Phi_{\mu\nu}=\partial_{\mu}\phi\partial_{\nu}\phi$) and the electromagnetic field (in which $\Phi_{\mu\nu}=F_{\mu}{}^{\alpha}F_{\alpha\nu}$).

\subsection{Special case}
In this section, we limit our analysis to the simplest form by setting $ \Phi^{\mu\nu} = u^{\mu} \,u^{\nu}$. In this case, the coefficients of the covariant and contravariant forms are related by
\begin{equation}
\nonumber
A=\fracc{1}{\alpha},\hspace{1cm} B = - \frac{\beta}{\alpha(\alpha+\beta)},
\end{equation}
where we set $u_{\mu}u_{\nu}\eta^{\mu\nu}=1$ and write the metric in the form
$$  \hat q^{\mu\nu} = \alpha \eta^{\mu\nu} + \beta \, u^{\mu} \,u^{\nu}.
$$

The associated covariant derivative is defined by
$$ u^{\alpha}{}_{|| \, \mu} = u^{\alpha}{}_{, \, \mu} + \widehat{\Gamma}^{\alpha}_{\mu\nu} \, u^{\nu}, $$
where the corresponding Christoffel symbol is constructed using $\hat q^{\mu\nu}$. The description of an accelerated curve\footnote{Note that we are dealing with a collection of paths $\Gamma$ that is usually called a congruence of curves. It is understood that each element of this collection concern particles that have the same characteristics. For instance, if the acceleration is due to an electromagnetic field, all particles of $\Gamma$ must have the same relation between its charge and mass, to wit a bunch of electrons.} in the flat space-time as a geodesics in the metric $\hat q_{\mu\nu}$
is possible if the following condition is satisfied
\begin{equation}
\left( u_{\mu , \nu} - \widehat{\Gamma}^{\epsilon}_{\mu\nu} \,
u_{\epsilon} \right) \, \widehat{u}^{\nu} = 0, \label{31dez01}
\end{equation}
where we have used the metric $\hat q^{\mu\nu}$ to write $\hat{u}^{\mu}\equiv \hat{q}^{\mu\nu} \, u_{\nu}=(\alpha+\beta)u^{\mu}$. Therefore,

\begin{equation}
\left( u_{\mu , \nu} - \widehat{\Gamma}^{\epsilon}_{\mu\nu} \,
u_{\epsilon} \right) \, u^{\nu} = 0.
\label{31dez1}
\end{equation}
In order to preserve the norm $\hat u^{\mu}\hat u_{\mu}$ along the curve we assume $\beta_{,\mu}u^{\mu}=0$, without loss of generality (it corresponds to a simple re-parametrization along the curves). Once the acceleration in the background is defined by $a_{\mu} = u_{\mu , \nu} \, u^{\nu} $, the condition of geodetic motion in the $\hat q_{\mu\nu}$-geometry takes the form

\begin{equation}
\label{31dez111}
a_{\mu} = \widehat{\Gamma}^{\epsilon}_{\mu\nu} \, u_{\epsilon} \, u^{\nu}.
\end{equation}
The Christoffel symbol reduces to
\begin{equation}
\widehat{\Gamma}^{\epsilon}_{\mu\nu} \, u_{\epsilon} \, u^{\nu} = \frac{\alpha + \beta}{2} \, u^{\alpha} \, u^{\nu} \, \widehat{q}_{\alpha\nu ,\mu}.
\label{31dez15}
\end{equation}
Using the expression of $ \widehat{q}_{\alpha\beta} $ in Eq.\ (\ref{31dez15}) and substituting the result into the condition (\ref{31dez111}), it follows that
$$ a_{\mu} =- \fracc{1}{2}\frac{\partial_{\mu}(\alpha+\beta)}{ (\alpha + \beta)}. $$
It means that the acceleration vector $a_{\mu}$ must be a gradient of a function $\Psi$, i.e.,

\begin{equation}
a_{\mu} \equiv \partial_{\mu}\Psi.
\label{31dez152}
\end{equation}
Thus, the expression of the coefficients $\alpha$ and $\beta$ of the metric $\hat q^{\mu\nu}$ are given in terms of the potential $\Psi$ of the acceleration by

\begin{equation}
\alpha + \beta = e^{- 2 \Psi}. \label{31dez155}
\end{equation}
This simple example gives a very useful formula, which exhibits the connection between geometrical and mechanical quantities. Later, in this paper, we shall analyze the case in which $\Psi$ represents the gravitational potential.

\subsection{Polynomial Metrics}
Gordon approach depends explicitly on the velocity $v^{\alpha}$ of the dielectric. Nevertheless such form of introducing an effective metric is not unique. Indeed it seems reasonable to present another metric $\hat q_{\mu\nu}$ that describes the same results obtained by Gordon and besides reduces the dependence on the four-velocity $v^{\alpha}$. From practical reasons, it might be useful to weaken this constraint of Gordon approach, once it is easier to determine the shape of the electromagnetic wave packet in the laboratory than constructing nonlinear dielectric media with arbitrary tensorial parameters $\epsilon_{\alpha\beta}$ and $\mu_{\alpha\beta}$---despite of the great advances in this research area recently \cite{smith,schuring}.

Let us now show that exists a class of geometries which play the same role as Gordon metric depending only on the angle $k_{\alpha}v^{\alpha}$ between the wave vector $k_{\alpha}$ and the dielectric four-vector $v^{\alpha}$. This is achieved by generalizations of the last section. Let us list some examples:\\

\underline{Case $A$}: the metric $\hat q^{\mu\nu}$ is given by

$$\hat q^{\mu\nu} = \alpha\eta^{\mu\nu} + \beta \, k^{\mu} \,k^{\nu},$$
and its inverse is
$$\hat q_{\mu\nu} = \fracc{1}{\alpha}\eta_{\mu\nu} - \fracc{\beta}{\alpha(\alpha+\beta N)} \, k_{\mu} \,k_{\nu}.$$

Once the wave vector $k_{\mu}$ is a gradient of a given hyper-surface $\Sigma$, then we have
$$k_{[\mu,\nu]}=0.$$
Substituting this result in Eq.\ (\ref{geo_gen}), it follows that $k_{\mu}$ must satisfy
$$\hat{N}_{(q),\mu}=0\hspace{.5cm}\Longrightarrow \hspace{.5cm} \hat{N}_{(q)}\equiv const.,$$
where we define $\hat N_{(q)}\equiv\hat q^{\mu\nu}k_{\mu}k_{\nu}$. That is, in order to follow a geodesic motion in $\hat q_{\mu\nu}$, $k_{\mu}$ must have constant norm in the metric $\hat q_{\mu\nu}$. The explicit expression for this constraint is

\begin{equation}
\label{mod_gord_k_q}
\hat{N}_{(q)}=(\alpha+\beta N)N\equiv1.
\end{equation}

Note that this approach transforms the non-normalized wave vector $k_{\mu}$ in Minkowski background in a normalized time-like vector in $\widehat q_{\mu\nu}$. It does not violate Lorentz invariance, because everything happens inside the dielectric. We note that it is not possible to fix $\hat N$ equal to zero, otherwise the metric is ill-defined. Therefore, $k_{\mu}$ is not a null-like vector in the $\hat Q$-metric. Another feature is that the magnitude of the dielectric four-vector
$$\hat q^{\mu\nu}v_{\mu}v_{\nu}=\alpha+\beta(k.v)^2,$$
is not necessarily positive definite allowing observers with velocity great than speed of light inside the medium\footnote{In the laboratory, the angle between these two vectors is easier to manipulate than the dielectric velocity field only. We expect that this fact could be of reasonable utility in the research of analog models.}. For instance, if we set
$$\alpha+\beta(k.v)^2=0,$$
then, using Eq.\ (\ref{mod_gord_k}), we obtain
$$\beta=\fracc{(\mu_0\epsilon-1)\alpha}{N}.$$
Substituting this result in Eq.\ (\ref{mod_gord_k_q}), yields
$$\alpha=\fracc{1}{\mu_0\epsilon N}.$$

Therefore, the metric $\hat q^{\mu\nu}$ with these values of $\alpha$ and $\beta$ produces the following outcome: the wave vector $k_{\mu}$ becomes a normalized and time-like vector, while the dielectric velocity $v^{\mu}$, which was a time-like vector in the Minkowski background, becomes a null geodesic in $\hat q_{\mu\nu}$. Therefore, the causal structure is no more determined by $k_{\mu}$.\\

Remark that the metric $\hat q^{\mu\nu}$ presented in the precedent sections is not unique. We can enlarge the set of metrics that have the same properties showed above adding other terms to $\hat q^{\mu\nu}$ provided the condition\ (\ref{2}) is valid. To exemplify these cases we consider:\\

\underline{Case $B$}: the polynomial metric is given by

$$\hat m^{\mu\nu} = \eta^{\mu\nu} + \beta \, k^{\mu} \,k^{\nu}+ \delta a^{\mu}a^{\nu}.$$
It is straightforward to show that its inverse has an extra term
$$\hat m_{\mu\nu} = \eta_{\mu\nu} + B \, k_{\mu} \,k_{\nu}+ \Delta a_{\mu}a_{\nu}+ \Lambda  a_{(\mu}k_{\nu)},$$
where $(\,)$ means symmetrization. The coefficients of the inverse metric are
$$B = - \fracc{\beta( 1 - \delta a^2 )}{X},$$
$$ \Delta = -\fracc{\delta(1+\beta N)}{X},$$
and
$$ \Lambda = \fracc{\beta\delta \dot N}{2X}.$$
Here, we defined $a^2\equiv-a^{\alpha}a_{\alpha}$, $\dot N\equiv N_{,\mu}k^{\mu}$ and
$$X=1-\delta a^2+\beta N-\beta\delta\left(\fracc{\dot N^2}{4}+Na^2\right).$$

The appearance of an extra term also happens with the inverse metric when we consider instead of $a^{\mu}a^{\nu}$ a term of the form $a^{(\mu}k^{\nu)}$. In both cases an extra term is necessary breaking the polynomial symmetry between the metric and its inverse. Nevertheless we will present the calculations for this case focusing only on the metric containing the term $a^{\mu}a^{\nu}$ and indicating that the results are very similar when the other term is considered separately.

The geodetic motion condition for the wave vector leads to

$$\hat{N}_{(m)}=(1+\beta N)N+\fracc{\delta}{4}\dot N^2=0.$$

Note that this approach permits a null geodesic motion for the wave vector $k_{\mu}$. This is the simplest case in which we regain the main Gordon result ($k_{\mu}$ as a null geodesic). The sign of the norm of $v^{\mu}$ is undetermined and may be chosen equal to zero, as we saw in the previous case.\\

\underline{Case $C$}: the most general case involving first order derivatives of $k_{\mu}$ occurs when the metric is expressed in the form\footnote{If we use higher derivatives of $k_{\mu}$ greater than that which appears in the dynamics, then the metric tensor $\hat q_{\mu\nu}$ is ill-defined.}

$$\hat n^{\mu\nu} = \alpha\,\eta^{\mu\nu} + \beta \, k^{\mu} \,k^{\nu}+ \delta a^{\mu}a^{\nu}+\lambda a^{(\mu}k^{\nu)}$$
and its inverse is
$$\hat n_{\mu\nu} = \fracc{1}{\alpha}\eta_{\mu\nu} + B \, k_{\mu} \,k_{\nu}+ \Delta a_{\mu}a_{\nu}+ \Lambda  a_{(\mu}k_{\nu)}.$$
The covariant metric coefficients are given by
$$B = - \fracc{\beta( \alpha - \delta a^2 )+\lambda a^2}{Z},$$
$$ \Delta = -\fracc{\delta(\alpha+\beta N)-\lambda^2N}{Z},$$
and
$$ \Lambda = -\fracc{\lambda(2\alpha+\dot N\lambda)-2\beta\delta \dot N}{2Z},$$
where {\small $$Z=\alpha\left[\alpha^2-\alpha\delta a^2+\alpha\beta N+\alpha\dot N\lambda-(\beta\delta-\lambda^2)\left(\fracc{\dot N^2}{4}+Na^2\right)\right].$$}

Once it involves more degrees of freedom, we can regain all outcomes presented before, but with different algebraic relations. In particular, the magnitude of the wave vector in $\hat n_{\mu\nu}$ is set

$$\hat{N}_{(n)}=(\alpha+\beta N+\lambda\dot N)N+\fracc{\delta}{4}\dot N^2.$$

Remark that the metric and its inverse have the same number of polynomial terms as required from the beginning. It did not happen in the case $B$ where an extra term was necessary in the inverse metric expression. Following this reasoning, in the next section we shall use only the cases $A$ and $C$ which satisfy the conditions (\ref{2612112}) and (\ref{2612111}). Moreover, as an example, we will set the potential $\Psi$ as being the Newtonian potential.

\subsection{Application: identifying $\Psi$ with the gravitational potential}

In the weak field limit the description of Newton's gravity can be formulated in terms of a geometric representation of the gravitational field. It can be done making use of effective potentials, which correspond to the well-known parameterized post-Newtonian approximation (PPN) \cite{will}. In this section, we compare some PPN results---particularly, those concerning general relativity (GR) predictions, which improve the Newtonian theory of gravity in the solar system---with some $\hat q-$metric given by the free-falling (geodesic) condition. We stress that $\hat q_{\mu\nu}$ just mimic the geodesic motion characteristic of the solutions of GR. Note that there is no dynamical equation for $\hat q_{\mu\nu}$ and we are not proposing such theory. It is purely a kinematical analogy.

This section shows that the analysis through geodesic paths are much more general than GR, once it is nothing but a choice of the metric. According to Poincar\'e's ideas upon geometrical descriptions of the world \cite{poincare}: \textit{non-Euclidean geometry is as legitimate as our ordinary Euclidean space; the enunciation of the Physics in this modified geometry would become more complicated, but it still would be possible.} In other words, it is possible to choose the metric of the space-time, such that an accelerated motion in a given geometry can be described as geodesic in another one.

For convenience, consider Minkowski space-time in spherical coordinates
$$ds^2=dt^2-dr^2-r^2d\Omega^2,$$
and an observer field $u_{\mu}=(1,f(r),0,0)$, which is a gradient $u_{\mu}\equiv\partial_{\mu}\Sigma$, where $\Sigma=t\pm F(r)$\footnote{This situation can perfectly be adapted to describe wave vectors in material media.}. This vector has a non-null acceleration given by

\begin{equation}
\nonumber
a_{\mu}=u_{\mu;\nu}u^{\nu}=\fracc{1}{2}N_{,\mu}=(0,-ff',0,0),
\end{equation}
where prime $'$ means derivative with respect to radial coordinate $r$ and $N\equiv u_{\mu} u_{\nu} \eta^{\mu\nu}$.

We choose the scalar function $\Psi$, which characterizes the acceleration $a_{\mu} \equiv \partial_{\mu}\Psi$, as identified with the Newtonian potential. Then, it follows a relation between $N$ and $\Psi$ given by

$$N=1+2\Psi=1-\fracc{r_H}{r},$$
where $r_H\equiv 2M$ and $M$ is the mass source of the gravitational field ($G=c=1$). To go further in the calculations, we basically split the analysis in two distinct cases. One of them gets immediately a wrong linear regime compared to GR, while the other case, which is separated in two subcases, can give the expected weak field regime.\\

\underline{Case I}:\, consider the $\hat q$-metric as follows

\begin{equation}
\label{drag_schw_v}
\hat q^{\mu\nu} =  \eta^{\mu\nu} + \beta \, u^{\mu} \,u^{\nu}.
\end{equation}
The inverse metric is

\begin{equation}
\label{inv_drag_schw_v}
\hat q_{\mu\nu} =  \eta_{\mu\nu} - \fracc{\beta}{1+\beta N} \, u_{\mu} \,u_{\nu}.
\end{equation}

The condition for $u_{\mu}$ to follow a geodesic motion in this metric is provided by

\begin{equation}
\label{no_norm_drag_cond}
u_{\mu||\nu}\hat u^{\nu} = \fracc{1}{2}\hat{N}_{(q),\mu}=0,
\end{equation}
where $\hat u^{\mu}\equiv\hat q^{\mu\nu}u_{\nu}=(1+\beta N)u^{\mu}$ and the magnitude of $u_{\mu}$ in $\hat q_{\mu\nu}$-metric is
$$\hat{N}_{(q)}\equiv\hat u^{\mu}u_{\mu}=(1+\beta N)N,$$
which is imposed by Eq.\ (\ref{inv_drag_schw_v}) to be constant different from zero. For convenience, we set $\hat{N}_{(q)}=1$.

A power law expansion in terms of $\epsilon\sim r_H/r$ of metric\ (\ref{inv_drag_schw_v}), which corresponds to the weak field limit, gives the following expressions

\begin{equation}
\label{q_schw_drag}
\begin{array}{lcl}
\hat q_{00}&\approx&1-\fracc{r_H}{r}\left(1+\fracc{r_H}{r}\right)+{\cal O}(\epsilon^3),\\[2ex]
\hat q_{01}&\approx&-\left(\fracc{r_H}{r}\right)^{3/2}\left(1+\fracc{r_H}{r}\right)+{\cal O}(\epsilon^{7/2}),\\[2ex]
\hat q_{11}&\approx&-1-\left(\fracc{r_H}{r}\right)^2\left(1+\fracc{r_H}{r}\right)+{\cal O}(\epsilon^4).\\[2ex]
\end{array}
\end{equation}
We see that this metric does not correspond to linearized Schwarzschild solution (even in Painlev\'e-Gullstrand coordinates due to the power $3/2$ instead of $1/2$ in $\hat q_{01}$). This metric is similar to some post-Newtonian approximation if we consider a rectilinear moving source for the gravitational field. The angular components of the metric are identical to Minkowski ones.\\

\underline{Case II}: let us consider another geometry $\hat n_{\mu\nu}$ given by

\begin{equation}
\label{drag_schw_g_v}
\hat n_{\mu\nu} =  \eta_{\mu\nu} + B\, u_{\mu} \,u_{\nu}+\Delta\, a_{\mu} \,a_{\nu}+\Lambda\, a_{(\mu} \,u_{\nu)}.
\end{equation}
The metric components are explicitly written as

\begin{equation}
\label{drag_schw_g_v_exp}
\begin{array}{lcl}
\hat n_{00}&=&1+B,\\[2ex]
\hat n_{01}&=&f(B-\Lambda f'),\\[2ex]
\hat n_{11}&=&-1+f^2(B+\Delta f'^2-2\Lambda f').\\[2ex]
\end{array}
\end{equation}
The other spatial components are identical to Minkowski metric in spherical coordinates. The condition which led $u_{\mu}$ to follow a geodesic motion in this metric is imposed on its magnitude

\begin{equation}
\label{norm_con_g}
\hat{N}_{(n)}=\fracc{N-\Delta(a^2N+\dot N^2/4)}{{\cal Z}}\equiv const.,
\end{equation}
where $\hat{N}_{(n)}\equiv \hat n_{\mu\nu}u^{\mu}u^{\nu}$. We also define

$${\cal Z}=\left(1+\fracc{\dot N\Lambda}{2}\right)^2-a^2\Delta+NB+a^2N\Lambda^2-B\Delta\left(a^2N+\fracc{\dot N^2}{4}\right).$$
In this case, we can set either $\hat{N}_{(n)}=0$ or $\hat{N}_{(n)}=1$. Let us analyze both cases separately:\\

\underline{Case II.a}: if $\hat{N}_{(n)}=0$, the assumption\ (\ref{norm_con_g}) implies that
$$\Delta=\fracc{N}{a^2N+\dot N^2/4}.$$

Once $u_{\mu}$ is null-like in $\hat n-$metric, we expect to regain the metric component of the PPN approximation responsible to describe correctly the light propagation deflection, which is given by the $\hat n_{11}$ component of the metric. So, using the considerations above, we set

\begin{equation}
\label{drag_schw_g_v_exp_n0}
\begin{array}{lcl}
\hat n_{00}&\approx&1+\fracc{r}{r_H}-\fracc{r_H}{r}+{\cal O}(\epsilon^2),\\[2ex]
\hat n_{01}&\approx&0,\\[2ex]
\hat n_{11}&\approx&-1-\fracc{r_H}{r}+\left(\fracc{r_H}{r}\right)^2+{\cal O}(\epsilon^3).\\[2ex]
\end{array}
\end{equation}
Note that a strange linear term appears in $\hat n_{00}$ due to our assumptions. If we try to avoid this term making some coordinate transformation, then the asymptotically flat regime is lost by other metric components, in such way that the trouble persists.\\

\underline{Case II.b}: in the case of $\hat{N}_{(n)}=1$, we can correctly reproduce the linear approximation of Schwarzschild solution. That is,

\begin{equation}
\label{drag_schw_g_v_exp_n1}
\begin{array}{lcl}
\hat n_{00}&\approx&1-\fracc{r_H}{r}+2\left(\fracc{r_H}{r}\right)^2+{\cal O}(\epsilon^3),\\[2ex]
\hat n_{01}&\approx&0,\\[2ex]
\hat n_{11}&\approx&-1-\fracc{r_H}{r}+{\cal O}(\epsilon^3).\\[2ex]
\end{array}
\end{equation}
Note that $\hat n_{00}$ differs from PPN results already in second order approximation in $\epsilon$ (whose the expected value in $\epsilon^2$ is $1/2$ instead of $2$). Therefore, it will surely produce some discrepancy in high order terms of the expansion.

Remark that these calculations was basically done in order to illustrate some analogies between these geometries and those studied in analog models of gravity. For this reason, the first order approximation is enough to show their strong correlation. If someone takes this approach looking for a perfect kinematical analogy between Schwarzschild metric and some $\hat n_{\mu\nu}$, then the introduction of a nonlinear scalar potential $\Psi$ in Newtonian equation of acceleration becomes necessary and, therefore, it can reproduce exactly the Schwarzschild geodesics, for instance. Notwithstanding, we will not enter into the details of this generalization because it involves a complicated question about the physical meaning of the dynamics of such non-linear potential whereas in this paper we want to discuss only kinematical properties of the particle trajectory.

\section{Conclusion}
In this paper, we presented an extension of Gordon metric constructing a larger class of geometries which describes accelerated motions in Minkowski space-time as geodesics in an effective geometry $\hat q_{\mu\nu}$. This effective metric depends only on the background metric and on the velocity vector of the accelerated body. In particular, we analyzed accelerated paths of light inside a moving dielectric and constructed a class of geometries with peculiar properties in comparison to Gordon's approach, but with similar kinematical effects. Ultimately, we gather this new class of geometries, that we call $\hat Q$-metrics, with the effective geometries of nonlinear electromagnetism, producing a collection of possible metric structures of the space-time, which has several applications in the theory of analog models of gravity. We will come back to this in the future.

\section*{APPENDIX: Hadamard's Method for Discontinuities}

We analyze the discontinuities of the electromagnetic field according to the standard Hadamard method and obtain the dispersion relation for the wave vector $k_{\mu}$. Let $\Sigma$ be a surface of discontinuity of the field $A_{\mu}$. The discontinuity of an arbitrary function $f$ is given by:
\begin{equation}
\label{DiscDef}
{\left[ f(x) \right]}_\Sigma = \lim_{\epsilon \to 0^+} \bigl( f(x +
\epsilon) - f(x - \epsilon)\bigr).
\end{equation}
The field $A_{\mu}$ and its first derivative $\partial_\nu A_{\mu}$ are
continuous across $\Sigma$, while the second derivatives present a
discontinuity:
\begin{eqnarray}\label{CondHadamard}
\left[ A_{\mu} \right]_\Sigma &=& 0,\\
\left[ \partial_\nu A_{\mu} \right]_\Sigma &=& 0,\\
\left[ \partial_\alpha\partial_\beta A_{\mu} \right]_\Sigma &=& k_\alpha k_\beta
\xi_{\mu}(x),
\end{eqnarray}
where $ k_\mu \equiv \partial_\mu \Sigma$ is the propagation vector and
$\xi_{\mu}(x)$ is the amplitude of the discontinuity. Substituting these
discontinuity properties in the equation of motion
$$\eta^{\alpha\nu}F_{\mu\nu;\alpha}\equiv\eta^{\alpha\nu}A_{[\mu,\nu];\alpha}=0,$$ it follows that:
\begin{equation*}
k_\alpha k_\beta \eta^{\alpha\beta}=0.
\end{equation*}
This means that the discontinuities of the electromagnetic field propagate as null geodesics in the Minkowski metric $\eta_{\mu\nu}$.

\section{acknowledgements}
We acknowledge J. M. Salim, J. D. Toniato, E. Goulart, V. de Lorenci and R. Klippert for their comments and the staff of ICRANet in Pescara where this work was partially done. We would like to thank FINEP, FAPERJ, CNPq and CFC/CBPF for their financial support.


\begin{thebibliography}{50}
\bibitem{gordon}
W. Gordon, {\em Ann.\ Phys.\ (Leipzig)} {\bf 72} 421  (1923); F. W. Hehl and Y. N. Obukhov, \textit{Foundations of classical electrodynamics: charge, flux, and, metric}, Progress in Mathematical Physics, v. 33, Birkh\"auser (2003);
\bibitem{novello1}
M. Novello, V. A. De Lorenci, J. M. Salim and R. Klippert {\em Phys.\ Rev.\ D} {\bf 61} 045001 (2000) and references therein;
\bibitem{pendry}
J. B. Pendry et al., {\em Science} {\bf 312} 1780 (2006);
\bibitem{landau}
L. Landau and E. Lifshitz, \textit{Electrodynamique des milieux Continus}, Ed. Mir, Moscow (1969);
\bibitem{hadamard}
J. Hadamard, \textit{Le\c cons sur la propagation des ondes et les \'equations de hydrodynamique}, Ed. Hermann, Paris (1903); V. D. Zakharov, \textit{Gravitational waves in Einstein's theory}, John Wiley \& Sons, Inc., New York (1973);
\bibitem{novello2}
M. Novello and E. Goulart {\em Class.\ Quantum\ Grav.} {\bf 28} 145022 (2011) and references therein;
\bibitem{novello3}
M. Novello, S. E. Perez-Bergliaffa and J. M. Salim, {\em Phys.\ Rev.\ D} {\bf 69} 127301 (2004);
\bibitem{novello4}
M. Novello, S. E. Perez-Bergliaffa, J. M. Salim, V. A. De Lorenci and R. Klippert, {\em Class.\ Quantum\ Grav.} {\bf 20} 859 (2003);
\bibitem{hawk_rad}
S. M. Hawking, {em Nature} {\bf 248} 30 (1974);
\bibitem{unruh}
W. G. Unruh, {\em Phys.\ Rev.\ Lett.} {\bf 46} 1351 (1981);
\bibitem{nov_viss_volo}
M. Novello, M. Visser and G. Volovik (Eds), \textit{``Artificial Black Holes", Proceedings of the workshop: Analog Models of General Relativity}, World Scientific (2002);
\bibitem{barcelo}
C. Barcel\'o, S. Liberati and M. Visser, {\em Liv.\ Rev.\ Rel.} {\bf 14} 3 (2011);
\bibitem{feyn}
R. Feynman et al., \textit{Feynman Lectures on Gravitation}, Westview Press, 1 ed., (1995);
\bibitem{smith}
D. R. Smith et al., {\em Science} {\bf 305} 788 (2009);
\bibitem{schuring}
D. Scuring et al., {\em Science} {\bf 314} 977 (2006);
\bibitem{will}
C. Will, {\em Liv.\ Rev.\ Rel.} {\bf 9} 3 (2006)
\bibitem{poincare}
H. Poincar\'e, \textit{Science and Hypothesis}, Ed. Scott, Michigan University (1905).
\end{thebibliography}
\end{document}